\documentclass[]{aastex701}

\newcommand{\logg}{\ensuremath{\log(g)}} 

\newcommand{\mum}{$\mu$m~}

\usepackage{CJK}
\usepackage{wrapfig}

\turnoffediting

\begin{document}
\begin{CJK}{UTF8}{gbsn}


\title{From Hubble to HWO: Bridging the Frontier of White Dwarf Exoplanet Science}

\suppressAffiliations

\author[orcid=0000-0002-3553-9474,sname='L. K. Rogers']{Lead Authors: Laura K. Rogers}
\altaffiliation{Affiliations listed at the end of the manuscript}
\affiliation{NOIRLab, 950 N Cherry Ave, Tucson, AZ, 85719, USA}
\email[show]{laura.rogers@noirlab.edu}  

\author[orcid=0000-0002-8808-4282]{Siyi Xu (许\CJKfamily{bsmi}偲\CJKfamily{gbsn}艺)} 
\affiliation{NOIRLab, 950 N Cherry Ave, Tucson, AZ, 85719, USA}
\email{siyi.xu@noirlab.edu}

\author[orcid=0000-0002-7116-3259]{Co-authors: Martin Barstow} 
\affiliation{School of Physics and Astronomy, University of Leicester, University Road, Leicester LE1 7RH, UK}
\email{martin.barstow@leicester.ac.uk}

\author[orcid=0000-0002-9632-1436]{Simon Blouin} 
\affiliation{Department of Physics and Astronomy, University of Victoria, Victoria, BC V8W 2Y2, Canada}
\email{sblouin@uvic.ca}

\author[orcid=0000-0002-8070-1901]{Amy Bonsor} 
\affiliation{Institute of Astronomy, University of Cambridge, Madingley Road, Cambridge CB3 0HA, UK}
\email{abonsor@ast.cam.ac.uk}

\author[orcid=0000-0003-0105-5540]{Andrew M. Buchan} 
\affiliation{Department of Physics, University of Warwick, Coventry, CV4 7AL, UK}
\email{Andy.Buchan@warwick.ac.uk}

\author[orcid=0000-0003-2478-0120]{Sarah L. Casewell} 
\affiliation{School of Physics and Astronomy, University of Leicester, University Road, Leicester LE1 7RH, UK}
\email{slc25@leicester.ac.uk}

\author[orcid=0000-0001-7296-3533]{Tim Cunningham} 
\affiliation{Center for Astrophysics | Harvard \& Smithsonian, 60 Garden St, Cambridge, MA 02138, USA}
\email{timothy.cunningham@cfa.harvard.edu}

\author[orcid=0000-0002-1783-8817]{John Debes} 
\affiliation{Space Telescope Science Institute, Baltimore, MD, USA}
\email{debes@stsci.edu}

\author[orcid=0000-0003-4609-4500]{Patrick Dufour} 
\affiliation{Trottier Institute for Research on Exoplanets, D\'epartement de physique, Universit\'e de Montr\'eal, 1375 Ave. Th\'er\`ese-Lavoie-Roux Montr\'eal, QC H2V 0B3, Canada}
\affiliation{Centre de recherche en astrophysique du Qu\'ebec, Montr\'eal, QC, Canada}
\affiliation{Observatoire du Mont-M\'egantic, Montr\'eal, QC, Canada}
\email{patrick.dufour@umontreal.ca}

\author[orcid=0000-0002-2761-3005]{Boris G\"{a}nsicke} 
\affiliation{Department of Physics, University of Warwick, Coventry, CV4 7AL, UK}
\email{boris.gaensicke@gmail.com}

\author[orcid=0000-0001-9632-7347]{Joseph Guidry} 
\affiliation{Department of Astronomy \& Institute for Astrophysical Research Boston University Boston, MA 02215}
\email{jaguidry@bu.edu}

\author[orcid=0000-0002-5775-2866]{Ted von Hippel} 
\affiliation{Department of Physical Sciences, Embry-Riddle Aeronautical University, Daytona Beach, FL 32114, USA}
\email{ted.vonhippel@erau.edu}

\author[orcid=0000-0001-6098-2235]{Mukremin Kilic} 
\affiliation{Homer L. Dodge Department of Physics and Astronomy, University of Oklahoma, 440 W. Brooks Street, Norman, OK 73019, USA}
\email{kilic@ou.edu}

\author[orcid=0000-0002-3307-1062]{\'Erika Le Bourdais} 
\affiliation{Trottier Institute for Research on Exoplanets, D\'epartement de physique, Universit\'e de Montr\'eal, 1375 Ave. Th\'er\`ese-Lavoie-Roux Montr\'eal, QC H2V 0B3, Canada}
\email{erika.le.bourdais@umontreal.ca}
\affiliation{Centre de recherche en astrophysique du Qu\'ebec, Montr\'eal, QC, Canada}
\affiliation{Observatoire du Mont-M\'egantic, Montr\'eal, QC, Canada}

\author[orcid=0000-0001-9834-7579]{Carl Melis} 
\affiliation{Department of Astronomy and Astrophysics, University of California, San Diego, La Jolla, CA 92093-0424, USA}
\email{cmelis@ucsd.edu}

\author[orcid=0009-0002-6065-3292]{Lou Baya Ould Rouis} 
\affiliation{Department of Astronomy \& Institute for Astrophysical Research Boston University Boston, MA 02215}
\email{lbor@bu.edu}

\author[orcid=]{Judith Provencal} 
\affiliation{Department of Physics and Astronomy and SARA, University of Delaware, Newark, DE 19716, USA}
\email{jlprov@gmail.com}

\author[orcid=0000-0001-7493-7419]{
Melinda Soares-Furtado} 
\affiliation{ Department of Astronomy, University of Wisconsin–Madison, 475 N. Charter Street, Madison, WI 53706, USA}
\affiliation{Wisconsin Center for Origins Research, University of Wisconsin–Madison, 475 N Charter Street, Madison, WI 53706, USA}
\affiliation{ Department of Physics, University of Wisconsin–Madison, 1150 University Avenue, Madison, WI 53706, USA}
\email{msoares.physics@gmail.com}

\author[orcid=0000-0001-6515-9854]{Andrew Swan} 
\affiliation{Department of Physics, University of Warwick, Coventry, CV4 7AL, UK}
\email{andrew.swan@warwick.ac.uk}

\author[orcid=0000-0002-4872-1021]{Isabella Trierweiler} 
\affiliation{ Department of Astronomy, Yale University, New Haven, CT 06511, USA}
\email{isabella.trierweiler@yale.edu}

\author[orcid=0000-0002-0853-3464]{Zachary Vanderbosch} 
\affiliation{Hobby-Eberly Telescope, 32 Mt. Locke Rd., McDonald Observatory, TX 79734, USA}
\email{zachary.vanderbosch@austin.utexas.edu}

\author[orcid=0009-0007-8709-9689]{Jamie Williams} 
\affiliation{Department of Physics, University of Warwick, Coventry, CV4 7AL, UK}
\email{Jamie.T.Williams@warwick.ac.uk}




\section{Executive Summary}

White dwarf stars, the endpoint of stellar evolution for 97\% of stars in our Milky Way, offer a unique and powerful window into the bulk elemental composition of rocky exoplanetary bodies. Up to 50\% of single white dwarfs are observed with photospheric metal lines from accreted exoplanetary bodies (called `polluted' white dwarfs), and spectroscopic observations reveal the bulk composition of this material. High-resolution ($R\,>15,000$) UV spectra are \textit{essential} for detecting many elements present in the material, such as the volatile elements imperative for habitability studies (C, N, O, P, S) and key rock-forming elements required to constrain interior structure (e.g. Fe, Si, Mg, Al, Ni). \textit{HST}, through its COS and STIS spectrographs, remains the only facility capable of performing this science in the near future. Looking to the next decade, the scientific case for continued \textit{HST} UV observations of polluted white dwarfs is compelling on three fronts (i) as a standalone to enable the bulk composition of exoplanetary material to be measured in a statistically significant sample, (ii) as essential groundwork for the \textit{Habitable Worlds Observatory} (\textit{HWO}), and (iii) in a powerful synergy with \textit{JWST}, to enable characterization of the bulk mineralogy \textit{and} bulk elemental composition of exoplanetary material. This white paper argues that continued UV spectroscopic capabilities with \textit{HST} is a high-return investment for white dwarf and exoplanet science, and preserving and prioritizing \textit{HST}'s UV capabilities through at least 2035 is crucial to maximize the scientific return from \textit{HST}, \textit{JWST}, and \textit{HWO}.

\section{\textit{HST} as a tool to study rocky exoplanet composition using white dwarfs} \label{pollution}

\begin{figure}[h]
    \centering \includegraphics[width=1.0\columnwidth,keepaspectratio=true]{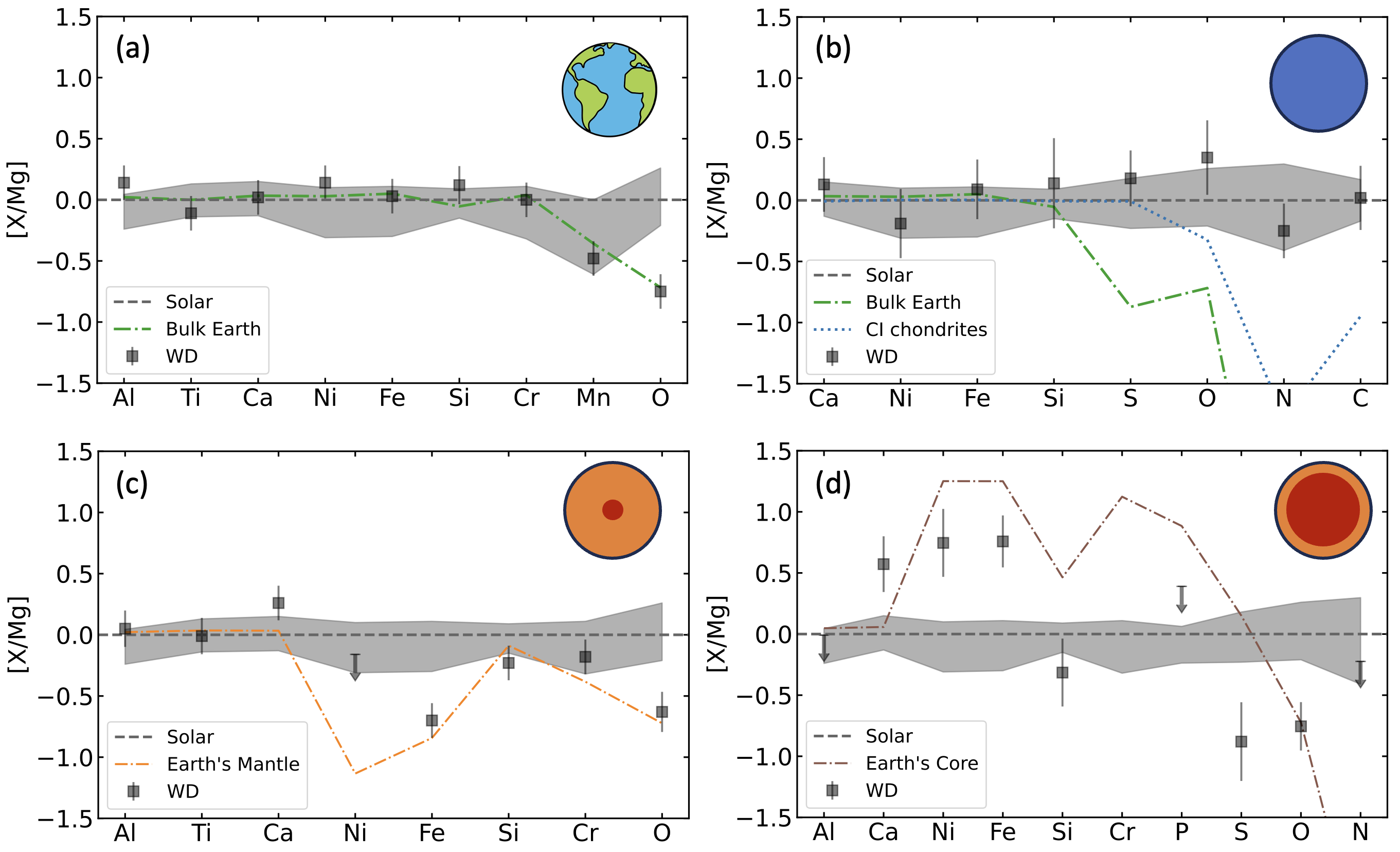}
    \caption{The logarithmic number ratios of elemental abundances (ordered in increasing volatility/decreasing condensation temperature) relative to Mg for the planetary material accreting onto four white dwarfs, normalized to solar composition (black dashed line at $[\mathrm{X/Mg}]=0$) for (a) Earth-like material \citep{Jura2015evidence}, (b) icy water-rich material \citep{xu2017chemical}, (c) mantle-rich material \citep{Rogers2023sevenI,Rogers2024sevenII}, (d) core-rich material \citep{Williams2025measurements}. The gray region shows the abundance ratios for main sequence FG stars \citep{hinkel2014stellar}, and the lines show the abundances of bulk Earth, CI chondrites, Earth's mantle, and Earth's core (assuming 90\% core and 10\% mantle) for comparison \citep{mcdonough2003compositional,Lodders2009solar,Lodders2025solar}. With the exception of Ca and Ti, which are only detectable optically, all these diagnostic elements require \textit{HST} UV spectroscopy for high-precision measurements.} \vspace{-3mm}
    \label{fig:compositions}
\end{figure}

The discovery of more than 6,000 confirmed exoplanets has revealed extraordinary diversity in planetary demographics, yet a fundamental limitation persists: whilst we can measure exoplanet masses, radii, and orbital properties, this only enables educated guesses about their interior structure and bulk elemental/mineralogical composition. Without this, we cannot meaningfully assess geological evolution, volatile inventories, or the potential to support life, questions which motivate the next big missions such as Ariel, PLATO, \textit{Roman} and the  \textit{HWO}. Polluted white dwarfs offer a direct solution to this problem, and \textit{HST} is the only facility that can unlock it at scale. 

White dwarfs are the final evolutionary stage for $\sim$97\% of all stars, including our Sun and most observed exoplanet host stars. Their extreme surface gravities ($\logg\approx8$) cause heavy elements to gravitationally settle below the photosphere on timescales as short as days--years, leaving a pure hydrogen or helium photosphere. The detection of heavy metals for between 25--50\% of white dwarfs therefore signals ongoing accretion of tidally disrupted planetary bodies \citep[e.g.][]{zuckerman2003metal,koester2014frequency,OuldRouis2024Constraints}. Measuring the abundances of these metals from optical and UV spectra enables the determination of the exoplanetary material's bulk elemental composition. Measuring the full suite of diagnostic elements (C, N, O, P, S, Mg, Si, Fe, Al, Ni, etc.) requires the UV ($\lambda < 3000$\AA), a spectral window accessible only from space and uniquely provided by \textit{HST}'s COS and STIS spectrographs. The scientific returns from \textit{HST} UV spectra of polluted white dwarfs over the past two decades have been nothing less than transformative \citep[e.g.][]{Xu2024chemistry}, but data collection has been piecemeal and exploratory. The next decade offers the opportunity to advance from individual case studies to systematic, statistically powerful, volume-complete surveys. Below we summarize key science outcomes already enabled by \textit{HST}, and the step-change in understanding that an extended program would enable. 

\textbf{Are rocky, Earth-like compositions the norm?} The majority of polluted white dwarfs studied to date exhibit refractory elemental abundance patterns consistent with dry, rocky compositions, analogous to bulk Earth and chondritic meteorites \citep[e.g.][]{gansicke2012chemical,Jura2015evidence,doyle2023new,Trierweiler2023chondritic,Xu2024chemistry}. Volatile elements are systematically depleted relative to Solar ratios (an example is illustrated in Figure \ref{fig:compositions}a), mirroring the condensation and volatile-loss processes that shaped the terrestrial planets of the Solar System. These signatures, appearing across diverse white dwarf systems, demonstrate that the formation of dry, rocky bodies is a near-universal outcome of planetary system evolution. However, these conclusions are limited by small number statistics and the fact that few statistically unbiased studies exist. An extended \textit{HST} mission would establish the true distribution of exoplanetary bulk compositions, quantify how common Earth-like rocky material is across planetary systems, and provide definitive statistical conclusions that isolated case studies cannot deliver.

\textbf{What is the volatile inventory of exoplanetary material?} A subset of white dwarfs display high oxygen abundances that cannot be explained by rocky oxides alone, implying the accretion of water-rich exoplanetary material 
\citep[see Figure \ref{fig:compositions}b; e.g.,][]{farihi2011possible, Rogers2024sevenII, Trierweiler2025water}. These were the first direct detections of extrasolar water-rich bodies, demonstrating that icy planetesimals survive the violent post-main-sequence phase, directly evidencing the pathways for water retention and delivery central to sustaining habitable environments. Beyond water, two systems show evidence for the accretion of Kuiper-belt like material with N- and C-rich ices \citep{xu2017chemical,Sahu2025Discovery}, opening a window into the full volatile architecture of exoplanetary material. Volatile ratios are a key prediction of protoplanetary disk models, which could be tested with such data \citep{Drazkowska2026dust}. Most C, N, O, P, and S detections require UV spectroscopy, however, only three polluted white dwarfs have the full suite of detections. An extended \textit{HST} mission would expand this, enabling population-level characterization of volatile abundances, the chemical forms in which they are carried, and their relationship to the architectures of the parent planetary systems. These measurements carry direct implications for understanding the prevalence of the conditions necessary for habitable environments across the Galaxy. 

\textbf{Is core--mantle differentiation widespread beyond the Solar System?} If a planetesimal is sufficiently hot and/or massive it will melt and segregate into an iron-rich core and an iron-poor mantle in a process called differentiation. This is crucial for planetary evolution as it controls magnetic dynamos, tectonic activity, and long-term habitability. White dwarfs that accrete fragments of disrupted, differentiated bodies reveal the chemical fingerprints of their internal compositions \citep{buchan2022planets}: mantle-dominated accretion produces lithophilic (iron-depleted) abundance patterns (e.g. Figure \ref{fig:compositions}c), whilst core-rich accretion yields iron and siderophilic enhancements (e.g. Figure \ref{fig:compositions}d). UV spectroscopy is essential for measuring the full complement of siderophilic and lithophilic elements that distinguish these scenarios, with current evidence suggesting that differentiation is common outside the Solar System \citep[e.g.][]{bonsor2020exoplanetesimals}. An extended \textit{HST} mission would enable population-level characterization of core--mantle differentiation, moving the field from isolated case studies to statistically robust insights into the processes that govern the fragmentation and the ultimate fate of differentiated bodies \citep{Buchan2024white}. Additionally, the abundances of Si, Ni and Cr (most often detected in the UV) can constrain the conditions under which differentiation occurs. In particular, these metals are sensitive to the mass of the parent body and oxygen fugacity, both of which have strong implications for the habitability of these worlds \citep{buchan2022planets}.

\begin{wrapfigure}{r}{0.5\textwidth}
    \centering
    \includegraphics[width=0.43\textwidth, keepaspectratio=true]{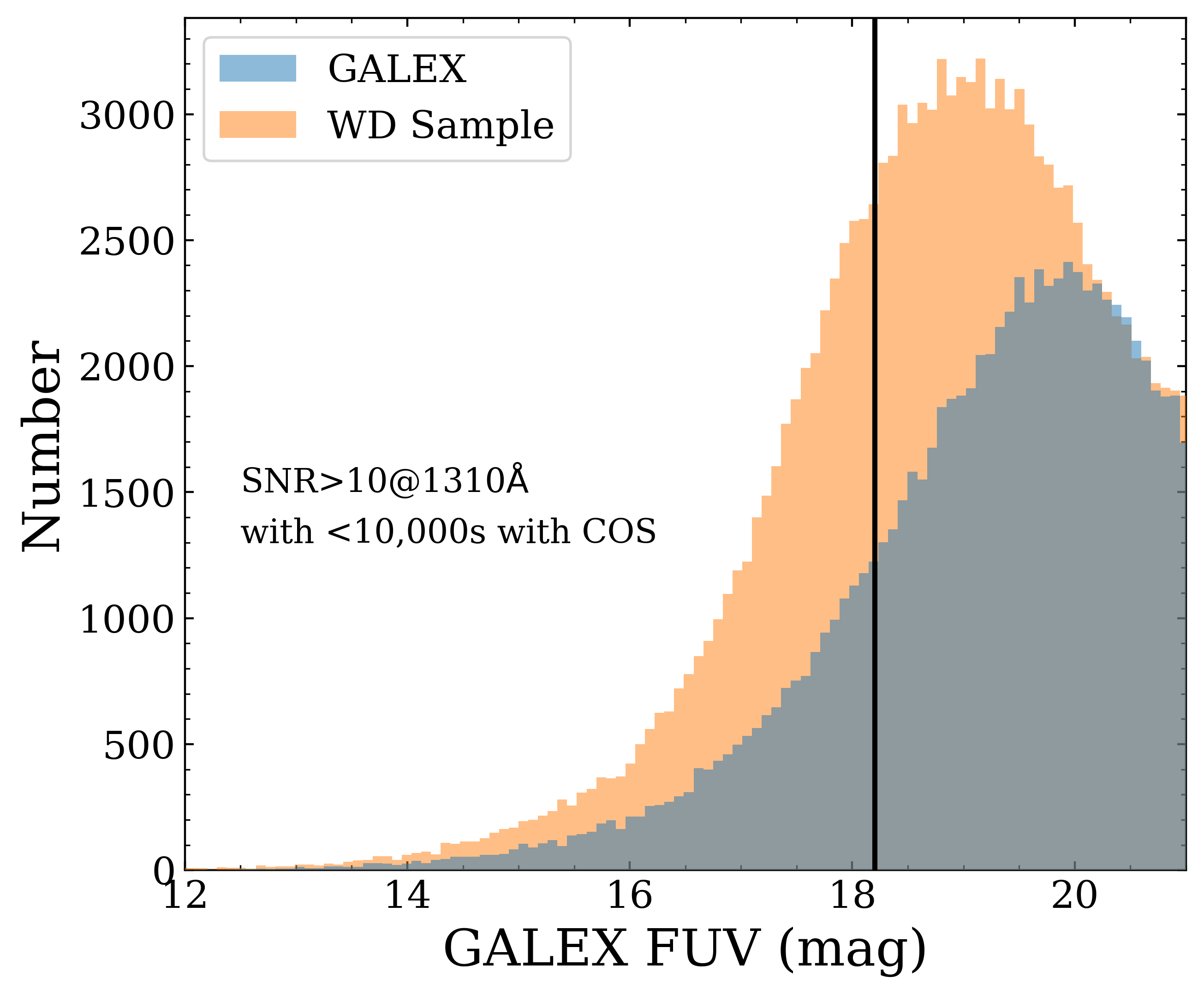}
    \caption{Number of white dwarfs as a function of GALEX FUV magnitude. The blue histogram represents the existing GALEX detections \citep{Wall2023GALEX}, while the orange distribution highlights the expected \textit{UVEX} detections from the white dwarf sample \citep{Gentile2021catalogue}. The black line marks the magnitude limit of COS/130M at 1310~{\AA}, where a signal to noise ratio (SNR) of at least 10 can be obtained with less than 10,000~sec ($\approx$ 5 orbits).}
    \label{fig:GALEX}
\end{wrapfigure}

\textbf{Characterizing the full diversity of extrasolar planet compositions, as outlined above, requires a large sample of polluted white dwarfs -- especially those with a large number of elements from refractories to volatiles.} \textit{Gaia} EDR3 has revealed  $\sim$359,000 white dwarfs \citep{Gentile2021catalogue}, but \textit{HST} remains the only facility capable of obtaining the UV spectroscopy needed to characterize their pollution, with the most efficient strategy being targeted follow-up observations of heavily polluted systems already identified from large-scale multi-object spectroscopic surveys. The supply of such targets is growing rapidly: Dark Energy Spectroscopic Survey Instrument (DESI) Data Release 1 alone has identified around 1000 new polluted white dwarfs \citep[][submitted]{Swan2026}, with 4-meter Multi-Object Spectroscopic Telescope (4MOST, \citealt{Chiappini2019}), Sloan Digital Sky Survey (SDSS)-V \citep{SDSSV}, and William Herschel Telescope Enhanced Area Velocity Explorer \citep[WEAVE, ][]{Jin2024WEAVE} set to expand coverage across both hemispheres. An extended mission is essential to exploit the influx of targets coming over the next decade. 

The UV Explorer (UVEX) mission will survey the entire sky at a depth 50 and 100 times deeper than \textit{GALEX} in the near- and far-UV, respectively \citep{Kulkarni2021UVEX}. Notably, UVEX does not have a bright source limit, enabling it to survey the Galactic Plane -- a region not observed by \textit{GALEX} -- thereby significantly expanding the UV coverage and identifying missing white dwarfs. The expected UVEX performance relative to existing \textit{GALEX} detections is shown in Figure~\ref{fig:GALEX}, and UVEX will likely detect more than 20,000 bright white dwarfs suitable for \textit{HST} followups. In addition, UVEX's low-resolution spectroscopy will facilitate the identification of the most heavily polluted white dwarfs for a more detailed characterization with \textit{HST}. UVEX is scheduled to launch in 2030 with a two-year primary mission lifetime. Furthermore, the Canadian Space Agency's CASTOR mission will also identify many polluted white dwarfs with the low-resolution UV spectroscopy in the 2030s \citep{Cote2025CASTOR}. Therefore, it is essential to keep \textit{HST} until at least 2035 to take full advantage of new discoveries from UVEX and CASTOR.

\section{Synergy with {\em HWO}}

The \textit{HWO} is one of NASA's next flagship missions. It will have a high-resolution UV spectrograph and 6 times the collecting area of \textit{HST}, possibly with wavelength coverage down to 900~{\AA} -- a range essential for detecting volatile elements such as nitrogen, which are key to understanding the volatile content of extrasolar planetary material \citep{Xu2026_HWO}. Current \textit{HST} observations of polluted white dwarfs are therefore critical to curate the most promising targets for future observations with \textit{HWO}. Without an extended \textit{HST} mission, \textit{HWO} will survey a handful of well-characterized systems, spending its early years on discovery rather than probing the underlying physics.

\section{Synergy with {\em JWST}}

The same process that pollutes white dwarfs also produces an observable circumstellar disk. When a planetesimal is perturbed onto a star-grazing orbit, tidal forces disrupt it at the Roche limit, resulting in a close-in dust disk with temperatures of $\sim$\,800--1200\,K \citep{jura2003tidally}. This thermal emission produces a characteristic infrared excess detectable at wavelengths beyond $\sim2$\,$\mu$m, with more than 100 such disks already confirmed in the pre-\textit{JWST} era \citep[e.g.][]{lai2021infrared}. Critically, infrared spectroscopy of these disks reveals their \textit{mineral} composition \citep[e.g.][]{reach2009dust,Reach2025GD362}, one of few methods that enable direct measurements of exoplanetary minerals. 

Mineralogy greatly influences planetary habitability. Whilst bulk elemental composition sets the ingredient list, mineralogy determines how those ingredients are assembled and how they behave geophysically. The dominant rock-forming silicates (olivine, pyroxene, quartz, silica) are built from the key rock-forming elements: Mg, Si, Fe, and O, which make up more than 90\% of bulk Earth \citep{mcdonough2003compositional}. The relative abundances (e.g. Mg/Si ratio) and crystal structures (e.g. Fe content) of these phases govern a planet's capacity to store water in its mantle, drive plate tectonics and volcanic activity, and sustain the geochemical cycling that underpins the long-term habitability of terrestrial planets \citep{Kelley2010mantle,Rogers2025Silicate}. Therefore, measuring the mineralogy of exoplanetary material is crucial for understanding habitability and whether rocky planets around other stars could sustain life, and this is possible with white dwarf disks. 

\begin{figure}[h]
    \centering \includegraphics[width=0.91\columnwidth,keepaspectratio=true]{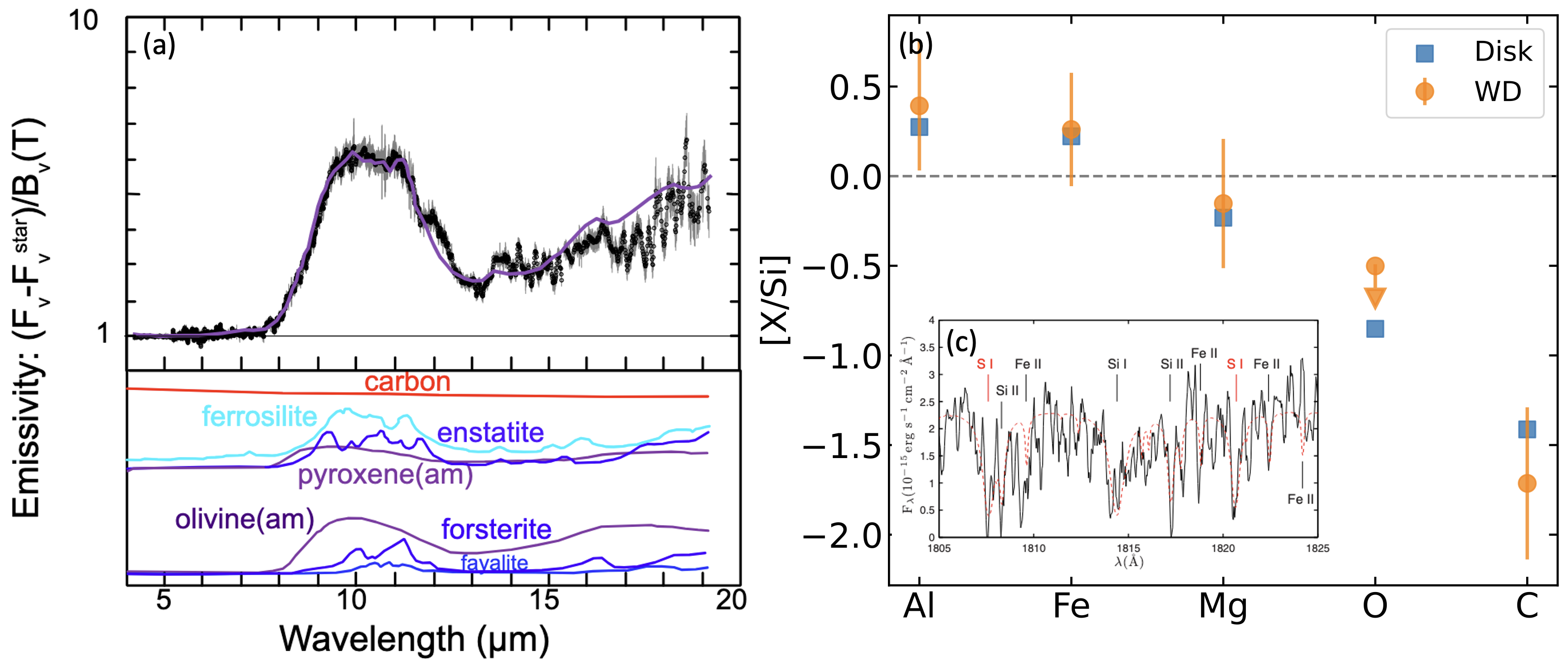}
    \caption{(a) \textit{JWST} MIRI Spectrum of the disk around GD\,362 from \citet{Reach2025GD362} with a model fit showing the minerals contributing to the emissivity. (b) Comparison of the composition of the planetary material accreting onto the white dwarf with that derived from the disk. (insert, c) \textit{HST} spectra of GD\,362 showing lines from S, Si and Fe from \citet{xu2013two}. This highlights how \textit{HST} UV spectroscopy reveals bulk composition of exoplanetary material and \textit{JWST} reveals its mineralogy enabling a direct comparison.} \vspace{-1mm}
    \label{fig:GD362}
\end{figure}

\textit{Spitzer} observed eight white dwarf disks \citep{reach2005dust,jura2009six,Rogers2025Silicate}, but it could only provide detailed mineralogical constraints for the brightest system. \textit{JWST} has sensitivity orders of magnitude beyond \textit{Spitzer} and wider infrared wavelength coverage with the NIRSpec and MIRI spectrographs. \textit{JWST} is now detecting and characterizing dust around white dwarfs in unprecedented detail \citep[][see Figure \ref{fig:GD362}a]{Farihi2025subtle,Reach2025GD362}, with the 10\,\mum silicate emission feature (seen in Figure \ref{fig:GD362}a) being diagnostic of mineral phases, and grain size distributions. The coming decade will see \textit{JWST} build a sample of dozens of characterized white dwarf disks, enabling population-level studies of disk mineralogy and evolution for the first time. 

The combination of \textit{JWST} infrared disk spectroscopy and \textit{HST} UV photospheric abundance measurements of the same exoplanetary material is uniquely powerful. Figure \ref{fig:GD362}b demonstrates this directly: the mineralogy derived from the disk of GD\,362 shows remarkable consistency with the bulk elemental abundances measured from its photosphere \citep{Reach2025GD362}, with further works also demonstrating this connection \citep{xu2018infrared,Rogers2025Silicate,Okuya2026dirtiness}. Together, \textit{HST} and \textit{JWST} deliver the full geochemical fingerprint of exoplanetary material: disk mineralogy anchored by bulk elemental composition, enabling self-consistent integrated chemical-mineralogical models of exoplanetary material. This enables evaluations of planetary properties such as volcanism, crustal recycling, and long-term nutrient supply, all key factors for sustained planetary habitability. The results will establish a new benchmark for linking exoplanetary geochemistry and the conditions necessary for life. This joint analysis is only now becoming possible with \textit{JWST}, but it requires \textit{HST} UV spectroscopy of every disk-bearing system that \textit{JWST} characterizes. This is only possible with an extended \textit{HST} mission as \textit{JWST} will continue to discover and characterize new systems beyond the current projected lifetime of \textit{HST}.

Variability has been observed across all components of polluted white dwarf systems: in circumstellar gas disks, dust disks, photospheric metal absorption lines, and the disintegrating planetesimals feeding the disk \citep[e.g.][]{wilson2014variable,vanderburg2015disintegrating,Noor2025Activity,Farihi2026accretion}. However, the physical mechanisms connecting these components remain poorly understood. Key open questions include: how material is transported from the circumstellar reservoir to the white dwarf surface, what sets the viscous timescales governing gas disk evolution, how dust is produced, destroyed, and replenished, and how the interplay between accretion rate variability and gravitational settling shapes the observed photospheric abundances at any given epoch, with implications for gas and dust interactions across vast areas of astrophysics. Answering these questions requires simultaneous, time-resolved measurements across the full spectral baseline from \textit{HST} UV spectra, where photospheric metal lines and circumstellar gas lines trace instantaneous accretion rate and the circumstellar reservoir, to the infrared where \textit{JWST} traces the dust reservoir. In turn, this enables measurements of how changes in disk structure and gas content propagate into photospheric abundance variations and directly constrain the viscous timescales and the lag between disruption and collision events and their photospheric signatures. Furthermore, there are $\sim$20 white dwarfs exhibiting transient or recurring photometric dimming by tidally disrupted planetesimal debris \citep{Bhattacharjee2025ZTF}, with Rubin/LSST set to increase this number drastically. These events can last just a few weeks and evolve on timescales as short as hours, therefore rapid \textit{HST} and \textit{JWST} follow-up would enable tidal disruption events to be studied in real time, capturing the onset of accretion and the earliest stages of disk formation. These are new and wide-open areas of study that require an extended \textit{HST} mission.

\vspace{-5pt}
\section{Requirements}

The science cases presented in this white paper require \textit{HST}'s UV spectroscopic capabilities to extend into the next decade and boosting \textit{HST} is essential to deliver this. The primary spectrograph for polluted white dwarf science is COS, which provides high UV sensitivity for the majority of targets. For hot white dwarfs ($>15,000$\,K), the diagnostic elements fall in the FUV, with COS/G130M (central wavelengths 1222 and 1291\,\AA) most commonly employed. For cooler white dwarfs ($<15,000$\,K), the NUV becomes essential, with COS/G185M and STIS/E140M, E230M, G230L being used. STIS is used less frequently than COS, but is crucial for observing bright white dwarfs where COS saturates. Calibration stability is equally critical as multi-epoch metal line monitoring requires reproducibility at the few percent level between visits spanning months to years. Identifying photospheric lines and disentangling them from the ISM also requires accurate relative and absolute radial velocity accuracy. The resolution of WFC3/UVIS is not sufficient to resolve lines and disentangle line contributions as we require $R\,>15,000$ for the science case. Beyond standard queue scheduling, rapid response access to follow up rapidly varying tidally disrupted bodies requires more Target of Opportunity or Director's Discretionary scheduling on timescales of days to weeks. Finally, we advocate for relaxing the standard 90\% program completeness threshold for statistical surveys that are sensitive to incompleteness. Unlike programs where a failed visit can be tolerated at a marginal loss, unbiased samples require completeness, and failed visits must be repeated. 

\textit{HST} is crucial for white dwarf planetary system science and its loss before \textit{HWO} would leave a critical gap in our ability to characterize the composition of exoplanet material. Key science questions, such as: what is the bulk composition and mineralogy of rocky exoplanets, how common is water in exoplanetary systems, how common are Earth-like compositions, and how frequent is core-mantle differentiation, all require UV spectroscopy below 3000\,\AA, a spectral window that COS and STIS uniquely probe at high-resolution. Answering these questions demands that \textit{HST} maintain its sensitive FUV and NUV spectroscopic capabilities through at least 2035.

\newpage
\bibliography{main}
\bibliographystyle{aasjournalv7}



\restoreAffiliations
\allauthors

\end{CJK}

\end{document}